\documentclass[twocolumn,preprintnumbers,amsmath,amssymb]{revtex4}

\usepackage{dcolumn}
\usepackage{bm}

\usepackage{graphicx}
\usepackage{verbatim}

\begin{document}

\title{Proton Decay and Fermion Masses in Supersymmetric SO(10) Model with Unified Higgs Sector}

\author{Yunfei Wu}

\author{Da-Xin Zhang}
 \email{dxzhang@phy.pku.edu.cn}
\affiliation{
School of Physics and State Key Laboratory of Nuclear Physics and Technology, Peking University,  Beijing 100871, China \\
}


\begin{abstract}
We make a detailed analysis on the proton decay in a supersymmetric SO(10) model proposed by K.Babu, I.Gogoladze, P.Nath, and R. Syed. 
We introduce quark mixing, 
and find that this model can generate 
fermion mass without breaking the experimental bound on proton decay. 
We also predict large CKM unitarity violations. 
\end{abstract}
\maketitle

\section{Introduction}
In grand unification models\cite{Georgi,Fritzsch} quarks and leptons are usually contained in the same multiplets. Consequently, baryon and lepton numbers are not conserved in general. If the models are supersymmetric, dimension-five operators mediated by the color-triplet Higgs superfields are dominant in these baryon and lepton number non-conservation processes\cite{sakai,Weinberg}. These dimension-five operators are also related to the fermion masses, thus are highly predictable in most of the supersymmetric unification models.

To build up unification models that generate correct fermion masses
and fulfill the stability of baryons, one usually needs to add into
more Higgs multiplets and/or more fermion multiplets. In
Ref.\cite{Babu}, a unified Higgs sector of $144 + \overline{144}$ is
used in the supersymmetric SO(10) model, so that the model is rather simple. The
fermion masses of the third generation are generated through cubic
couplings at the price of introducing extra heavy matter superfields of 45 and
10 dimension multiplets.  The fermions masses of the first two
generations arise from the Planck scale effects. The dimension-five
operators, which mediate proton decay, are not very tightly
connected to the fermion masses. Consequently, both the fermions
masses and the stable baryons are achieved in the model.

In our work, we analyze the problems of fermion masses and proton decay in this new model. By introducing quarks mixing, we obtain 
the CKM unitarity breaking effects and
the diagonal quark masses.
The general form of the dimension-five operators for
baryon decay is given. We pick out 
the most important proton decay mode $p\to K^+\bar\nu_\mu$.
We fit the quartic and cubic Yukawa couplings of the second
and third generation at unification scale. By using these values, we
get the couplings of dimension-five operators to analyze the proton lifetime.

This paper is organized in the following way: A short review of the
model is presented in section \ref{sec:sb}. In section
\ref{sec:mix}, we give the quark mixing and mass generation. The
CKM unitarity breaking is analyzed. In section \ref{sec:bl}, we
present the general dimension-five operators and low energy
Lagrangian for proton decay.
In section\ref{sec:num}, the numerical results are discussed. We found the
model can survive in some parameter space.
Finally, we summarize our results.

\section{\label{sec:sb}Review of the Model}
The model of  Ref.\cite{Babu} uses the following superpotential
\begin{eqnarray}
 {W}&=& M {(\overline{144}_H\times 144_H)}\nonumber\\
 &&+ \frac{\lambda_{45_1}}{M'} { (\overline{144}_H\times
144_H)_{45_1}
 (\overline{144}_H\times 144_H)_{45_1}}\nonumber\\
 &&+  \frac{\lambda_{45_2}}{M'} { (\overline{144}_H\times
144_H)_{45_2}
 (\overline{144}_H\times 144_H)_{45_2}}\nonumber\\
 &&+  \frac{\lambda_{210}}{M'} { (\overline{144}_H\times
144_H)_{210}
 (\overline{144}_H\times 144_H)_{210}},\label{Higgs}
\end{eqnarray}
where $M'$ is supposed to be at the Plank scale and $\lambda$'s are the couplings after integrating out the corresponding 45 or 210 dimension component fields. $M$ is the mass of 144 Higgs. Eq.(\ref{Higgs}) gives
the one step breaking of SO(10) to the supersymmetric stand model
and the doublet-triplet splitting.

The terms responsible for the symmetry breaking is
\begin{eqnarray}
{ W}_{_{SB}}=M{ \bf Q}^i_{j}{\bf P}^j_{i}
+\frac{1}{M'}\left[-\lambda_{45_{_1}}+\frac{1}{6}\lambda_{210}
\right] {
\bf Q}^i_{j}{ \bf P}^j_{i} {\bf Q}^k_{l}{ \bf P}^l_{k}\nonumber\\
+\frac{1}{M'}\left[-4\lambda_{45_{_1}}-\frac{1}{2}\lambda_{45_{_2}}-\lambda_{210}
\right] {\bf Q}^i_{k}{ \bf P}^k_{j} {\bf Q}^j_{l}{ \bf P}^l_{i},
\end{eqnarray}
where the all the fields are chiral supermultiplets and indices $i,
j$ take 1 to 5. $P$'s and $Q$'s are 24 dimension Higgs of SU(5) coming from 144 and $\overline {144}$ respectively.  To get the following vacuum expectation values and
minimization of ${ W}_{_{SB}}$
\begin{eqnarray}
 <{\bf Q}^i_{j}>&=& q~\textnormal{diag}(2,2,2,-3,-3),\nonumber\\
 <{\bf P}^i_{j}>&=& p~\textnormal{diag}(2,2,2,-3,-3),\label{mini}
\end{eqnarray}
we need
\begin{equation}\label{qp}
\frac{MM'}{qp}=116\lambda_{45_{_1}}+7\lambda_{45_{_2}}+4\lambda_{210}.
\end{equation}
The $D$-flat condition needs $q=p$. The vacuum expectation values in Eq.(\ref{mini}) break the SO(10) down to the Standard Model gauge group.

The further electroweak symmetry breaking requires two of the Higgs doublets
light and all the Higgs triplets heavy. The superpotential
governs the doublet-triplet splitting is given in
Ref.\cite{Babu,Babu1}. We
only give out masses splitting results. The Higgs doublet pairs and triplet pairs before
splitting are
\begin{eqnarray}
&{D}_1:({\bf Q}^{a}, {\bf P}_{a}),~{T}_1:({\bf Q}^{\alpha}, {\bf P}_{\alpha}),&\nonumber\\
&{D}_2:({\bf Q}_{a},{\bf P}^{a} ), ~{T}_2:({\bf Q}_{\alpha},{\bf P}^{\alpha}),&\nonumber\\
&{D}_3:({\bf{\widetilde Q}}_{a}, {\bf {\widetilde P}}^{ a
}), ~{T}_3:({\bf{\widetilde Q}}_{\alpha}, {\bf
{\widetilde P}}^{ \alpha
}),&\nonumber\\
&{T}_4:~({\bf{\widetilde Q}}^{\alpha}, {\bf {\widetilde
P}}_{ \alpha }),
\end{eqnarray}
where $\alpha, \beta$ take 1, 2 and 3 while $a, b$ take 4 and 5.
The diagonalization of the doublets' and triplets' mass matrices use the rotations 
\begin{eqnarray}\label{rotatedhiggs}
 \left[\begin{matrix}({\bf { Q}}^{\prime
}_{a},{\bf { P}}^{\prime a} )\cr ({\bf {\widetilde
Q}}^{\prime}_{a},{\bf {\widetilde P}}^{\prime a})
\end{matrix}\right]& =& \left[\begin{matrix} \cos\theta_{
D} & \sin\theta_{D}\cr -\sin\theta_{D} &
\cos\theta_{
D}\end{matrix}\right]\left[\begin{matrix}({\bf{ Q}}_{a}, {\bf {
P}}^{a})\cr ({\bf {\widetilde Q}}_{ a},{\bf {\widetilde P}}^{ a})
 \end{matrix}\right],\nonumber\\
 \left[\begin{matrix}({\bf { Q}}^{\prime
}_{\alpha},{\bf { P}}^{\prime \alpha} )\cr ({\bf {\widetilde
Q}}^{\prime}_{\alpha},{\bf {\widetilde P}}^{\prime \alpha})
\end{matrix}\right]& =& \left[\begin{matrix} \cos\theta_{T} &
\sin\theta_{T}\cr -\sin\theta_{T} &
\cos\theta_{
T}\end{matrix}\right]\left[\begin{matrix}({\bf{ Q}}_{\alpha}, {\bf
{ P}}^{\alpha})\cr ({\bf {\widetilde Q}}_{ \alpha},{\bf
{\widetilde P}}^{ \alpha})
 \end{matrix}\right],
\end{eqnarray}
where
\begin{eqnarray}\label{mix}
\tan\theta_{D}&=&\frac{1}{{d_3}}\left({
d_2}+\sqrt{{d_2}^2+{d_3}^2}\right),\nonumber\\
\tan\theta_{T}&=&\frac{1}{{t_3}}\left({
t_2}+\sqrt{{t_2}^2+{t_3}^2}\right).
\end{eqnarray}
Here
\begin{eqnarray}
{
d}_1&=&-\frac{2}{5}M+\frac{qp}{M'}\left(\frac{296}{5}\lambda_{45_1}
-16\lambda_{45_2}-\frac{392}{15}\lambda_{210}\right),\nonumber\\
{
d}_2&=&-\frac{8}{5}M+\frac{qp}{M'}\left(-\frac{1036}{5}\lambda_{45_1}
+\frac{1}{2}\lambda_{45_2}+\frac{427}{15}\lambda_{210}\right),\nonumber\\
{
d_3}&=&2\sqrt{\frac{3}{5}}\frac{qp}{M'}\left(10\lambda_{45_1}
+\frac{5}{4}\lambda_{45_2}-\frac{5}{6}\lambda_{210}\right).
\end{eqnarray}
The mass eigenvalues are found to be
\begin{eqnarray}\label{doublets}
M_{{D}_1}&=&M+\frac{qp}{M'}(180\lambda_{45_1}+9\lambda_{45_2}-10\lambda_{210}),\nonumber\\
M_{{D}_2,{D}_3}&=&\frac{1}{2}\left({
d_1}\pm\sqrt{{d_2}^2+{d_3}^2}\right).
\end{eqnarray}
We will set $M_{D_3}$ to be small, and get
\begin{equation}
H_u={\bf { \widetilde P}}^{\prime}_{a}, \text{and} ~H_d={\bf \widetilde { Q}}^{\prime}_{a}.
\end{equation}

Moreover,  the triplet eigenstates' masses are
\begin{eqnarray}
M_{{
T}_1}&=&M+\frac{qp}{M'}(180\lambda_{45_1}+4\lambda_{45_2}-10\lambda_{210}),\nonumber\\
M_{{
T}_4}&=&-M+\frac{qp}{M'}(-84\lambda_{45_1}-4\lambda_{45_2}+2\lambda_{210}),\nonumber\\
M_{{T}_2,{T}_3}&=&\frac{1}{2}\left({
t_1}\pm\sqrt{{t_2}^2+{t_3}^2}\right),
\end{eqnarray}
where
\begin{eqnarray}
{
t}_1&=&-\frac{2}{5}M+\frac{qp}{M'}\left(\frac{576}{5}\lambda_{45_1}
-11\lambda_{45_2}-\frac{302}{15}\lambda_{210}\right),\nonumber\\
{
t}_2&=&-\frac{8}{5}M+\frac{qp}{M'}\left(-\frac{816}{5}\lambda_{45_1}
-2\lambda_{45_2}+\frac{212}{15}\lambda_{210}\right),\nonumber\\
{t_3}&=&\sqrt{5}\frac{qp}{M'}\left(8\lambda_{45_1}
+\lambda_{45_2}-\frac{2}{3}\lambda_{210}\right).
\end{eqnarray}

The authors of \cite{Babu1} introduce two kinds of couplings to gain
fermion masses. 
All the three generations achieve masses from the quartic couplings\cite{Babu,Babu1}
\begin{eqnarray}
~\left\{\zeta_{ij}^{^{(10)}}\right\}~\left(16_{i}\times16_{j}\right)_{10}\left(144\times144\right)_{10},\nonumber\\
~\left\{\xi_{ij}^{^{(10)}}\right\}~\left(16_{i}\times16_{j}\right)_{10}\left(\overline{144}\times \overline{144}\right)_{10},\nonumber\\
~\left\{\varrho_{ij}^{^{(126)}}\right\}~\left(16_{i}\times
16_{j}\right)_{\overline{126}} \left(144\times144\right)_{126},\nonumber\\
~\left\{\lambda_{ij}^{^{(45)}}\right\}~\left(16_{i}\times
\overline{144}\right)_{45} \left(16_{j}\times \overline{144}\right)_{45},\nonumber\\
~\left\{\zeta_{ij}^{^{(120)}}\right\}~\left(16_{i}\times
16_{j}\right)_{120}\left({144}\times{144}\right)_{120},\nonumber\\
~\left\{\xi_{ij}^{^{(120)}}\right\}~\left(16_{i}\times16_{j}\right)_{{120}}
\left(\overline{144}\times \overline{144}\right)_{120},\nonumber\\
~\left\{\lambda_{ij}^{^{(54)}}\right\}~\left(16_{i}\times\overline{144}\right)_{54} \left(16_{j}\times\overline{144}\right)_{54},\nonumber\\
~\left\{\lambda_{ij}^{^{(10)}}\right\}~\left(16_{i}\times144\right)_{10}\left(16_{j}\times144\right)_{10},\label{quarticcoup}
\end{eqnarray}
where $i, j$ are the generation indices. The extra 10-plet and
45-plet extra heavy matters couple only to the third generation
fermions through following superpotential\cite{Babu1}
\begin{eqnarray}\label{third1}
 {W}^{16\times \overline{144} \times {45}}&=&
 \frac{1}{2!}h^{(45)}<{\widehat\Psi}_{(+)}^{*}|B\Gamma_{[\mu}|{\widehat\Upsilon}_{(+)\nu]}> \widehat { { {\bf F}}}_{\mu\nu}^{(45)},\nonumber\\
{W}_{{{ mass}}}^{(45)}&=&m_F^{(45)}\widehat { {\bf
F}}_{\mu\nu}^{(45)}\widehat { {\bf
F}}_{\mu\nu}^{(45)},\label{extra}
\end{eqnarray}
and
\begin{eqnarray}\label{third2}
 {W}^{16\times {144} \times {10}}&=&
h^{(10)}<{\widehat\Psi}_{(+)}^{*}|B|{\widehat\Upsilon}_{(-)\mu}>
 \widehat { { {\bf F}}}_{\mu}^{(10)},\nonumber\\
{W}_{{{ mass}}}^{(10)}&=&m_F^{(10)}\widehat { {\bf
F}}_{\mu}^{(10)}\widehat { {\bf
F}}_{\mu}^{(10)},\label{extra1}
\end{eqnarray}
where ${\widehat\Psi}$ represents the 16 dimension fermion and ${\widehat\Upsilon}$ represents the 144 dimension Higgs.
We follow the authors of \cite{Babu} defining
\begin{equation}
f^{()}\equiv ih^{()},
\end{equation}
to get the real couplings.
From the above coupling forms, we  can get all the Yukawa couplings contributing to the fermion masses.

\section{\label{sec:mix} Quark mixing and Mass Generation}
The model of Ref.\cite{Babu} provides a mechanism of generating
fermion masses. From Eqs.(\ref{quarticcoup}), (\ref{third1}) and
(\ref{third2}), we can deduce all the couplings. 
They are put in Appendix \ref{app:DT}.

We assume the quartic parts
of mass matrix for up type quarks are already diagonalized to reduce
complexity. The main difference from Ref.\cite{Babu1} is
the authors of \cite{Babu1} neglect the fact that the quartic
coupling matrices might induce mixing between light quarks and the extra
45- and 10-plets.
In this new scenario the down quark mass matrix can be written as
follows
\begin{eqnarray}
&&M_{d}=\\
&&\begin{array}{cc}\begin{array}{c}
{}\\~d^c\\s^c\\{}^{16}b^c\\~\\ ~{}^{10}b^c\\~\\ ~{}^{45}b^c \end{array}
\begin{array}{c}d~~~~~s~~~~{}^{16}b~~~~~~~~{}^{45}b~~~~~~~~{}^{10}b\\
\left(\begin{array}{ccccc}
m_{11}&m_{12}&m_{13}&0&0\\
m_{21}&m_{22}&m_{23}&0&0\\
m_{31}&m_{32}&m_{33}&m_b''&m_D^{(10)}\\
~&~&~&~&~\\
0&0&m_b' &0&-2m_f^{(10)}\\
~&~&~&~&~\\
0&0&m_D^{(45)}&-2m_f^{(45)}&0\end{array}\right)\end{array}
\end{array}\nonumber,\label{dmatrix}
\end{eqnarray}

where \begin{eqnarray}
{m_b}'=\frac{1}{ {2}}f^{(10)}
\left(\frac{\langle{\bf Q}_5\rangle}{\sqrt{10}}+\frac{\langle\widetilde{\bf
Q}_5\rangle}{2\sqrt 3}\right),\nonumber\\{m_b}''=-2\sqrt 2f^{(45)}\langle{\bf P}_5\rangle,\nonumber\\
m_D^{(45)}=-2\sqrt 2f^{(45)} p,m_D^{(10)}=\sqrt
2f^{(10)} q,
\end{eqnarray}
and all the $m$'s in the upper-left block are from the quartic couplings.

Noting that the quartic components of the upper-left $3\times3$ part of $M_d$ is extremely small, the cubic components will receive little effect from the quartic couplings when diagonalizing. In this sense, we can diagonalize the cubic part first by taking $m_{33}\sim0$ as in \cite{Babu} , and then consider how the quartic couplings take effects. The matrices $U_{t,b}$ and $V_{t, b}$ can be found in  Appendix A, which is slightly different from Ref.\cite{Babu1}. After diagonalizing the cubic part, we get
\begin{eqnarray}
U_bM_dV_b^{\text T}=~~~~~~~~~~~~~~~~~~~~~~~~~~~~~~~~~~~~~~~~~~~~~~~~~~~~~~~~~\nonumber\\
\begin{pmatrix}
m_{11}&m_{12}&m_{13}\cos\theta_{V_b}&m_{13}\sin\theta_{V_b}&0\\
m_{21}&m_{22}&m_{23}\cos\theta_{V_b}&m_{23}\sin\theta_{V_b}&0\\
m_{31}\cos\theta_{U_b}&m_{32}\cos\theta_{U_b}&\lambda_1&0&0\\
m_{31}\sin\theta_{U_b}&m_{32}\sin\theta_{U_b}&0&0&\lambda_2\\
0&0&0&\lambda_3&0\end{pmatrix}\label{dmatrix2},\nonumber\\
\end{eqnarray}
where $\lambda_2$ and $\lambda_3$ are the eigenmasses of rotated extra heavy fermions.


For $\lambda_2$ and $\lambda_3$ are extremely large,  we can
diagonalize the light down-type quark mass matrix
\begin{equation}
m_d^{ij}=\begin{pmatrix}
m_{11}&m_{12}&m_{13}\cos\theta_{V_b}\\
m_{21}&m_{22}&m_{23}\cos\theta_{V_b}\\
m_{31}\cos\theta_{U_b}&m_{32}\cos\theta_{U_b}&\lambda_1\\
\end{pmatrix}.
\end{equation}
If we denote the Yukawa couplings $m_d^{i}=m_d^{ij}V'_{jk}\delta^k_i$, in the up quark diagonalized basis $V'_{jk}$
 is analogous to the CKM matrix.

The matrices diagonalizing $M_d$ are
\begin{eqnarray}\label{vd}
&&V_d^{\text T}=V_b^{\text T}\times V'_{ij}+{\mathcal O}(\frac{\lambda_1}{\lambda_2},\frac{\lambda_1}{\lambda_3})+\cdots \nonumber\\
&&=\begin{pmatrix}
V'_{ud}&V'_{us}&V'_{ub}&0&0\\
V'_{cd}&V'_{cs}&V'_{cb}&0&0\\
V'_{td}\cos\theta_{V_b}&V'_{ts}\cos\theta_{V_b}&V'_{tb}\cos\theta_{V_b}&\sin\theta_{V_b}&0\\
-V'_{td}\sin\theta_{V_b}&-V'_{ts}\sin\theta_{V_b}&-V'_{tb}\sin\theta_{V_b}&\cos\theta_{V_b}&0\\
0&0&0&0&1
\end{pmatrix}\nonumber\\
&&+{\mathcal O}(\frac{\lambda_1}{\lambda_2},\frac{\lambda_1}{\lambda_3})+\cdots,
\end{eqnarray}
and
\begin{equation}
U_d=U_b+{\mathcal O}(\frac{\lambda_1}{\lambda_2},\frac{\lambda_1}{\lambda_3})+\cdots.
\end{equation}
The upper-left $3\times3$ part of $V_d^{\text T}$ is just the transpose of the CKM matrix.
When taking the quartic coupling for the up quarks to be diagonalized,
the up quark mass matrix can be diagonalized easily by
\begin{equation}
V_u^{\text T}=V_t^{\text T}
~~\text {and}~~
U_u=U_t.
\end{equation}

The mass matrices for the charged leptons have the same structure as the down
quarks for they share the same Yukawa couplings.

Beside mass matrix diagonalization, the CKM unitarity violation can also derived From Eq.(\ref{vd}) easily.
\begin{eqnarray}\label{ckmb}
|V_{ud}|^2+|V_{cd}|^2+|V_{td}|^2&=&1-|V_{td}|^2\tan^2\theta_{V_b},\nonumber\\
|V_{us}|^2+|V_{cs}|^2+|V_{ts}|^2&=&1-|V_{ts}|^2\tan^2\theta_{V_b},\nonumber\\
|V_{ub}|^2+|V_{cb}|^2+|V_{tb}|^2&=&1-|V_{tb}|^2\tan^2\theta_{V_b},\nonumber\\
|V_{td}|^2+|V_{ts}|^2+|V_{tb}|^2&=&1-\sin^2\theta_{V_b}.\end{eqnarray}
For $V_{td}$ and $V_{ts}$ are very small in the right-hand side of first two equations, the last two equations give the most important unitarity violation.
 We denote
\begin{eqnarray}
\delta_b&=&|V_{tb}|^2\tan^2\theta_{V_b},\nonumber\\
\delta_t&=&\sin^2\theta_{V_b}.
\end{eqnarray}

We need 
$|\tan\theta_{V_b}|\ll 1$ by
examing the unitarity bound on the CKM matrix\cite{PDG}. This is
different from Ref.\cite{nath}, where the authors took
$|\tan\theta_{V_b}|\gg 1$.  Under this new condition, the $b-\tau$
unification $f_b=f_\tau$ gives
\begin{equation}
\tan\theta_D=\frac{5\sqrt{30}}{83}.
\end{equation}
From section \ref{sec:num}, we will see that $\delta_t$ and
$\delta_b$ are given at the percent level by fixing the mass Yukawa
couplings.

The other CKM violations are
\begin{eqnarray}
V_{ud}V_{us}+V_{cd}V_{cs}+V_{td}V_{ts}&=&-V_{td}V_{ts}\tan^2\theta_{V_b},\nonumber\\
V_{us}V_{ub}+V_{cs}V_{cb}+V_{ts}V_{tb}&=&-V_{ts}V_{tb}\tan^2\theta_{V_b},\nonumber\\
V_{ud}V_{ub}+V_{cd}V_{cb}+V_{td}V_{tb}&=&-V_{td}V_{tb}\tan^2\theta_{V_b}.
\end{eqnarray}
They are related to the phenomena of flavor changing neutral currents, which are not presently concerned.

\section{\label{sec:bl} Dimension-Five Operators and Decay Rates}
In supersymmetric unification models, the dominant mechanism of
inducing proton decay is through the color-triplet Higgsino
mediation. The resulting dimension-five operators are of the type of
LLLL and RRRR. We will focus on LLLL-type only to simplify our
discussion althouth the RRRR-types can also be important\cite{Goto}.

The Yukawa coupling of the Higgs to the matter multiplets are as follows
\begin{eqnarray}
W_{Y}&=&h_u^iu_i^cQ_iH_u-V^*_{ij}f_d^jd_i^cQ_jH_d-f_e^ie_i^cL_iH_d\nonumber\\
&&+Y_{Qf}^iQ_iQ_iH_{cf}+Y_{Lf}^{ij}Q_iL_j{\bar H_{cf}}\nonumber\\
&&+Y_{e^cf}^{ij}u^c_ie^c_jH_{cf}+Y_{q^cf}^{ij}u^c_id^c_j{\bar H_{cf}},
\end{eqnarray}
where $h_u^i$'s, $f_d^j$'s and $f_e^i$'s are the Yukawa couplings giving masses and
$Y$'s are the Yukawa coupling with the color triplet Higgs. $f$ denotes different color triplets.

The dimension-five operators that cause the nucleon decay can be written explicitly as
\begin{eqnarray}
W_5&=&\frac{1}{M_{Tf}}Y_{Qf}^iY_{Lf}^{kl}(Q_iQ_i)(Q_kL_l)\nonumber\\
&&+\frac{1}{M_{Tf}}Y_{e^cf}^{ij}Y_{cf}^{kl}(u^c_ie^c_j)(u^c_kd^c_l).
\end{eqnarray}
The total anti-symmetry in
color index requires $i\not=k$, which implies the dominant mode is $p\to
K\bar\nu$\cite{Muraya}.


Dressing of wino to dimension-five operators gives the triangle diagram factor\cite{Arno,Muraya}
\begin{eqnarray}
f(u,d)&=&\frac{M_2}{m_{\tilde u}^2-m_{\tilde d}^2}\left(\frac{m_{\tilde u}^2}{m_{\tilde u}^2-M_2^2}\ln\frac{m_{\tilde u}^2}{M_2^2}\right.\nonumber\\
&&-\left.\frac{m_{\tilde d}^2}{m_{\tilde
d}^2-M_2^2}\ln\frac{m_{\tilde d}^2}{M_2^2}\right),\label{triangle}
\end{eqnarray}
where $M_2$ is the wino masses.
The resulting four-fermion operators can be written as
\begin{widetext}
\begin{eqnarray}
\mathcal L&=&Y(ijk)A_S(i,j,k)A_L\epsilon_{\alpha\beta\gamma}\left[ (u_i^\alpha d_i^{\prime\beta}) (d_j^{\prime\gamma} \nu_k)(f(u_j,\,e_k) + f(u_i,\,d'_i))+ (d_i^{\prime\alpha} u_i^\beta) (u_j^\gamma e_k)(f(u_i,\,d_i) + f(d'_j,\,\nu_k))\right. \nonumber \\
& & \left.+ (d_i^{\prime\alpha} \nu_k) (d_i^{\prime\beta} u_j^\gamma)(f(u_i,\,e_k) + f(u_i,\,d'_j))+ (u_i^\alpha d_j^{\prime\beta}) (u_i^\gamma e_k)(f(d'_i,\,u_j) + f(d'_i,\,\nu_k)) \right],\label{lag1}
\end{eqnarray}
where the coupling $Y(ijk)$ defined as follows
\begin{eqnarray}
Y(ijk)&=&\Big\{\frac{1}{M_{T1}}Y_{Q1}^iY_{L1}^{jk}+\frac{V^*_{jk}}{M_{T2}}\Big[Y_{Q2}^iY_{L2}^k\cos^2\theta_T+Y_{Q3}^iY_{L3}^k\sin^2\theta_T+(Y_{Q3}^iY_{L2}^k+Y_{Q2}^iY_{L3}^k)\cos\theta_T\sin\theta_T\Big]\nonumber\\
&&+\frac{V^*_{jk}}{M_{T3}}\Big[Y_{Q2}^iY_{L2}^k\sin^2\theta_T+Y_{Q3}^iY_{L3}^k\cos^2\theta_T-(Y_{Q3}^iY_{L2}^k+Y_{Q2}^iY_{L3}^k)\cos\theta_T\sin\theta_T\Big]\Big\}\frac{\alpha_2}{2\pi}.
\end{eqnarray}
\end{widetext}

In Eq.(\ref{lag1}), the function $A_S$ refers to the short range renormalization effect between the unification and the
supersymmetry breaking scale and $A_L$ the long range renormalization effect between the supersymmetry scale and 1 GeV. All of these have been investigated thoroughly in \cite{Nano,Muraya}.

The relevant terms for $p\to K^++\bar\nu_\mu$ from Eq.(\ref{lag1}) are
\begin{eqnarray}\label{fl}
\mathcal L&=&A_L\epsilon_{\alpha\beta\gamma}\left((d^\alpha u^\beta) (s^\gamma \nu_\mu)+ (s^\alpha u^\beta)(d^\gamma \nu_\mu) \right)\nonumber\\
&&\times \left[ A_S(c,u,s) Y(2,1,2) V_{cs} V_{cd} (f(c,\,\mu) + f(c,\,d'))\right.\nonumber\\
&&\left.+ A_S(t,u,s) Y(3,1,2) V_{ts} V_{td} (f(t,\,\mu) + f(t,\,d'))\right].\nonumber\\
\end{eqnarray}

We neglect the $\nu_e$ mode for the smallness of the first generation Yukawa couplings 
The direct coupling to $\nu_\tau$ is suppressed by CKM matrix element.  Although the coupling to ${}^{10}\nu$ is order 1, $~{}^{10}\nu$ is heavy and its contribution to rotated $\nu_\tau$ is highly suppressed. 

Noting that $Y_{L1}^{ij}$ has only diagonal elements, it will have no contribution. So $\bar H_{c1}$ or the first term in $Y(i,j,k)$ does not contribute to Eq.(\ref{fl}).


We can use the chiral Lagrangian technique\cite{Claudson,Chadha} to obtain hadronic level matrix elements
\begin{equation}
\langle K^+|(u,d)_L s_L|p\rangle=\frac{\beta}{f}\left(1+\left(\frac{D}{3}+F\right)\frac{m_N}{m_B}\right),\label{had1}
\end{equation}
\begin{equation}
\langle K^+|(u,s)_L d_L|p\rangle=\frac{\beta}{f}\frac{2D}{3}\frac{m_N}{m_B},\label{had2}
\end{equation}
in the limit $m_{u,d,s}\ll m_{N,B}$. All the parameters can be found in \cite{Muraya,aoki}.

\section{\label{sec:num}Numerical Results and Discussion}
In this section, we present some numerical results. The recent
Super-Kamiokande bound on proton decay is\cite{PDG}
\begin{equation}
\tau_{p\to K^++\bar\nu}>1.6\times10^{33}\text{yrs}.
\end{equation}

In the present model, the doublet and triplet masses connect to the
144 Higgs mass. We keep a pair Higgs doublets light. 

When we setting $M'$ in Eq.(\ref{Higgs}) to be Planck scale
$10^{19}$GeV, we can get the relation between light Higgs masses
and 144 Higgs mass. This is a fine tuning problem relating to the
doublet-triplet splitting. We choose one of the three doublets to be
light while leaving others heavy. Then we automatically get the heavy triplet Higgs
masses.

From Eq.(\ref{Bangle}), the condition $\tan\theta_{U_b}\ll1$ and
$\tan\theta_{V_b}\ll1$ give
\begin{equation}
m_F^{(10)}\gg p,\text{and }~ m_F^{(10)}\gg q,
\end{equation}
if we take $f^{(10)}$ and $f^{(45)}$ to be of order 1.

The
unification scale can be
\begin{equation}
M_{GUT}\sim 2\times10^{16}\text{GeV}.
\end{equation}
The supersymmetry breaks at about 1 TeV.
All the sfermion masses used in Eq.(\ref{triangle}) are taken to be
 1TeV.
The wino mass is taken as  $M_2=300$GeV.

\begin{table}
\caption{\label{tab:proton}Proton life and CKM unitarity
violation.
$\delta_b$ is got with $|V_{tb}|=0.77$\cite{PDG}.}
\begin{ruledtabular}
\begin{tabular}{cccccc}
$\tan\beta$&2&3&6&10&20\\
$\zeta^{(10)}/10^{-20}\text{GeV}^{-1}$&-0.10&-0.20&-0.30&-0.40&-0.80\\
$\zeta^{(120)}/10^{-20}\text{GeV}^{-1}$&1.5&2.5&4.5&7.2&14.5\\
$\xi^{(10)}/10^{-20}\text{GeV}^{-1}$&-0.40&-0.10&-0.10&-0.70&-0.90\\
$\xi^{(120)}/10^{-20}\text{GeV}^{-1}$&5.0&2.8&2.3&6.2&7.2\\
$\lambda^{(45)}/10^{-19}\text{GeV}^{-1}$&-8.48&-1.58&-1.35&-7.82&-6.47\\
$\lambda^{(54)}/10^{-19}\text{GeV}^{-1}$&22.0&4.10&3.52&20.3&16.8\\
$f^{(45)}$&1.24&1.04&0.96&0.95&0.95\\
$f^{(10)}$&0.960&1.14&1.59&2.05&2.85\\
\hline $\tau_{p}/(10^{33}\text{yrs})$&1.4$\times10^2$&84&38&4.3&1.9\\
$\delta_b$/(\%)
&2.4&1.7&1.4&1.3&1.3\\
$\delta_t$/(\%)&3.0&2.2&2.0&1.8&1.8
\end{tabular}
\end{ruledtabular}
\end{table}
We take $p=q=10^{16}$GeV and $m_F^{(10)}=m_F^{(45)}=10^{17}$GeV. After
fine tuning $M_{D_3}$ to the order of $10^2$GeV, the other Higgs
doublet and triplets are fixed at the order of $10^{16}$GeV.
Because the number of  Yukawa couplings in this model are redundant, we can just choose some of them to
generate the light fermion masses. Here we take
$\lambda^{(10)}_{ij}=\varrho^{(126)}_{ij}=0$.
At the unification scale we fit
$\xi^{(10)}$, $\xi^{(120)}$, $\zeta^{(10)}$, $\zeta^{(120)}$, $f^{(10)}$ and $f^{(45)}$ to get the correct fermion masses.
Besides fermions masses, this model can have long enough proton
life time even without cancellation introduced in
Ref.\cite{nath}, which implies vanishing down-type fermion masses of the second generation. We get the longest proton life times without affecting the fermions masses by fine tuning $\lambda^{(45)}$ and taking $\lambda^{(54)}=-7/27 \lambda^{(45)}$.
 The results are given in table \ref{tab:proton}, while we take all the parameters in Eqs.(\ref{had1},\ref{had2}) as \cite{aoki}
\begin{eqnarray}
\beta=0.0118\text{GeV}^3,~~ D=0.8,~~ F=0.47,~~~~~~~ \nonumber\\f=0.131\text{GeV},~~
 m_N=0.94\text{GeV},~~ m_B=1.15.\text{GeV}.
\end{eqnarray}

From table \ref{tab:proton}, the longest possible proton life time
decreases when $\tan\beta$ increases. The present model dilute the
relation between fermion masses and dimension-five operators. We can
choose some parameters to suppress the dimension-five operators
without affecting the fermion masses. Even for very large
$\tan\beta$, this model could have long enough proton life time.

The unitarity breaking of the CKM matrix can be obtained directly
from Eq.(\ref{vd}). Unitarity is good for the first and second
columns of CKM up to ${\cal
O}(\frac{\lambda_1}{\lambda_2},\frac{\lambda_1}{\lambda_3})$,
because $\lambda_2$ and $\lambda_3$ are extra fermion masses
at about $10^{16}$GeV. 
For the first two equations of Eqs.(\ref{ckmb}), unitarity
violations are about $10^{-6}$ due to the smallness of $V_{td}$ and
$V_{ts}$, within experimental constraints\cite{PDG}. This model predicts relatively
large CKM breaking for $\delta_b$ and $\delta_t$. 
Their values are at percentage level.



\section{Summary}
In this work we analyze the supersymmetric SO(10) model with a
unified 144+$\overline{144}$ Higgs\cite{Babu,Babu1}. We introduce the fermion mixing, and find large unitarity violations of the CKM matrix. We find proton life time is in agreement with the experiment bound for a rather large value of  $\tan\beta$.

This work was supported in part by the National Natural Science
Foundation of China (NSFC) under the grant No. 10435040.

\appendix

\section{\label{app:DT}Mass Matrices and Yukawa Couplings}
In this section we list the detailed results of the mass
matrices and the diagonalization matrices. From Eqs.(\ref{quarticcoup}),(\ref{extra}) and (\ref{extra1})
we can get all the couplings for the component fields. We put all of them in tables \ref{tab:table1} and \ref{tab:table2}. The Baryon-Lepton number violating terms can also be found in Ref.\cite{nath}.
\begin{table*}
\caption{\label{tab:table1}All the quartic couplings that contribute to masses and Baryon-Lepton number violatings. Generation indices are neglected. All the couplings are matrices of the three generations.}
\begin{ruledtabular}
\begin{tabular}{lcccc}

 &&Coupling constants&Mass terms
&Baryon number violating\\ \hline
&$\widehat{\bf M}^{ij}\widehat
{\bf M}_{j}\widehat
{\bf P}^{k}_i \widehat
{\bf P}_k$&$2\left(-\lambda^{(45)}-\lambda^{(54)}+8\xi^{(10)}+\frac{8}{3}\xi^{(120)}\right)$&$-3p\left(Q^{a\alpha}d^c_\alpha+\epsilon^{ab}e^cL_b\right){\bf P}_a$&$2p\left(\epsilon^{\alpha\beta\gamma}u^c_\gamma d^c_\beta+Q^{\alpha b}L_{b}\right){\bf P}_\alpha$ \\

  \raisebox{2ex}[0pt]{$J_{1}$}&$\widehat{\bf M}^{ij}\widehat
{\bf M}_{k}\widehat
{\bf P}^{k}_j\widehat
{\bf P}_i$&$2\left(-\lambda^{(45)}+\lambda^{(54)}\right)$ &$\left(2p\,Q^{a\alpha}d^c_\alpha-3p\,\epsilon^{ab}e^cL_b\right){\bf P}_a$&$\left(2p\,\epsilon^{\alpha\beta\gamma}u^c_\gamma d^c_\beta-3p\,Q^{\alpha b}L_b\right){\bf P}_\alpha $\\
\hline
 &$\epsilon_{ijklm}\widehat{\bf M}^{ij}\widehat
{\bf M}^{kl}\widehat
{\bf P}^{m}_n\widehat
{\bf P}^n$&$-\frac{1}{\sqrt{5}}\xi^{(10)}$&$-24p\,u^c_\alpha Q^{\alpha b}\epsilon_{ba}{\bf P}^a$
 &$16p\left(u^c_\alpha e^+-\epsilon_{\alpha\beta\gamma}Q^{\beta}Q^\gamma\right){\bf P}^\alpha $\\
 \raisebox{2ex}[0pt]{$J_{2}$}&$\epsilon_{ijklm}\widehat{\bf M}^{in}\widehat
{\bf M}^{jk}\widehat
{\bf P}^{l}_n\widehat
{\bf P}^m$&$\frac{1}{\sqrt{5}}\left(-\lambda^{(45)}+\frac{4}{3}\xi^{(120)}\right)$&$6p\,u^c_\alpha Q^{\alpha b}\epsilon_{ba}{\bf P}^a$&$4p\left(\epsilon_{\alpha\beta\gamma}Q^{\beta}Q^\gamma-u^c_\alpha e^+\right){\bf P}^\alpha $\\\hline
 &$\epsilon_{ijklm}\widehat{\bf M}^{ij}\widehat
{\bf M}^{kl}\widehat
{\bf P}^{p}_n\widehat
{\bf P}^{nm}_{p}$&$2\xi^{(10)}$&$40p\,u^c_\alpha Q^{\alpha b}\epsilon_{ba}\widetilde{\bf P}^a$&$20p\left(u^c_\alpha e^+-\epsilon_{\alpha\beta\gamma}Q^{\beta}Q^\gamma\right)\widetilde{\bf P}^\alpha $\\
 $J_{3}$&$\epsilon_{ijklm}\widehat{\bf M}^{in}\widehat{\bf M}^{jk}\widehat{\bf P}^{p}_n\widehat{\bf P}^{lm}_{p}$&$-\frac{1}{2}\lambda^{(45)}-\frac{1}{2}\lambda^{(54)}+\frac{4}{3}\xi^{(120)}$&$-\frac{20}{3}p\,u^c_\alpha Q^{\alpha b}\epsilon_{ba}\widetilde{\bf P}^a$&$40p\left(u^c_\alpha e^+-\epsilon_{\alpha\beta\gamma}Q^{\beta}Q^\gamma\right)\widetilde{\bf P}^\alpha $\\
&$\epsilon_{ijklm}\widehat{\bf M}^{np}\widehat{\bf M}^{jk}\widehat{\bf P}^{k}_p\widehat{\bf P}^{lm}_{n}$&$\frac{1}{2}\left(\lambda^{(45)}-\lambda^{(54)}\right)$&$\frac{34}{3}p\,u^c_\alpha Q^{\alpha b}\epsilon_{ba}\widetilde{\bf P}^a$&$-20p\left(u^c_\alpha e^++\epsilon_{\alpha\beta\gamma}Q^{\beta}Q^\gamma\right)\widetilde{\bf P}^\alpha $\\
\hline
$K_{1}$&$\widehat{\bf M}^{ij}\widehat{\bf M}^{kl}\widehat{\bf Q}^{m}_n\widehat{\bf Q}^{n}$&$2\left(\frac{2}{15}\varrho^{(126)}+\zeta^{(10)}+\frac{2}{3}\zeta^{(120)}\right)$&$-24q\,u^c_\alpha Q^{\alpha b}\epsilon_{ba}{\bf Q}^a$&$16q\left(u^c_\alpha e^+-\epsilon_{\alpha\beta\gamma}Q^{\beta}Q^\gamma\right){\bf Q}_\alpha$\\\hline

&$\widehat{\bf M}^{ij}\widehat{\bf M}_j\widehat{\bf Q}^{k}_i\widehat{\bf Q}_{k}$&$\frac{1}{\sqrt{5}}\left(\frac{1}{15}\varrho^{(126)}+8\zeta^{(10)}+\frac{8}{3}\zeta^{(120)}\right)$&$-3q\left(Q^{a\alpha}d^c_\alpha+\epsilon^{ab}e^+L_b\right){\bf Q}_a$&$2q\,\left(-\epsilon^{\alpha\beta\gamma} u^c_\beta d^c_\gamma+Q^{\alpha b}L_b\right){\bf Q}_\alpha$\\
\raisebox{2ex}[0pt]{$K_{2}$}&$\widehat{\bf M}^{ij}\widehat{\bf M}_k\widehat{\bf Q}^{k}_i\widehat{\bf Q}_{j}$&$\frac{1}{\sqrt{5}}\left(\frac{4}{15}\varrho^{(126)}-\lambda^{(10)}+\frac{16}{3}\zeta^{(120)}\right)$&$\left(-2q\,Q^{a\alpha}d^c_\alpha+3q\,\epsilon^{ab}e^+L_b\right){\bf Q}_a$&$\left(2q\,\epsilon^{\alpha\beta\gamma} u^c_\beta d^c_\gamma+3q\,Q^{\alpha b}L_b\right){\bf Q}_\alpha$\\\hline

&$\widehat{\bf M}^{ij}\widehat{\bf M}_j\widehat{\bf Q}^{k}_l\widehat{\bf Q}^{l}_{ik}$&$2\left(\frac{1}{15}\varrho^{(126)}+8\zeta^{(10)}-\frac{8}{3}\zeta^{(120)}\right)$&$4q\left(Q^{a\alpha}d^c_\alpha+\epsilon^{ab}e^+L_b\right)\widetilde{\bf Q}_a$&$-5q\left(Q^{\alpha b}L_b-\epsilon^{\alpha\beta\gamma} u^c_\beta d^c_\gamma\right)\widetilde{\bf Q}_\alpha$\\

\raisebox{2ex}[0pt]{$K_{3}$}&$\widehat{\bf M}^{ij}\widehat{\bf M}_k\widehat{\bf Q}^{k}_l\widehat{\bf Q}^{l}_{ij}$&$\frac{4}{3}\left(-\frac{1}{5}\varrho^{(126)}+4\zeta^{(120)}\right)$&$\left(12q\,\epsilon^{ab}e^+L_b-\frac{4}{3}q\;Q^{a\alpha}d^c_\alpha\right)\widetilde{\bf Q}_a$&$\left(2q\,\epsilon^{\alpha\beta\gamma} u^c_\beta d^c_\gamma-3q\,Q^{\alpha b}L_b\right)\widetilde{\bf Q}_\alpha$\\

\end{tabular}
\end{ruledtabular}

\end{table*}

\begin{table*}
\caption{\label{tab:table2}All the cubic couplings that contribute to masses and Baryon-Lepton number violatings.}
\begin{ruledtabular}
\begin{tabular}{lcccc}
 &&Couplings&Mass terms&Baryon number violating\\ \hline

$J_1$&$\widehat{\bf M}_{i}\widehat
{\bf P}_{j}\widehat
{\bf F}^{ij}$&$-2\sqrt{2}f^{(45)}$&$\left(~{}^{16}b^c_\alpha ~{}^{45}Q^{\alpha a}+~{}^{16}L_b ~{}^{45}\tau^{c}\epsilon_{ba}\right){\bf P}_a$&$\left(-\epsilon^{\alpha\beta\gamma}~{}^{16}b^c_\beta ~{}^{45}t^{c}_\gamma+~{}^{16}L_a~{}^{45}Q^{a\alpha}\right){\bf P}_\alpha$ \\

$J_2$&$\epsilon_{ijklm}\widehat{\bf M}^{ij}\widehat
{\bf P}^k\widehat
{\bf F}^{lm}$&$\frac{1}{\sqrt{10}}f^{(45)}$&$4\left({}^{16}t^c_\alpha~{}^{45}Q^{\alpha b}\epsilon_{ba}+{}^{45}t^c_\alpha~{}^{16}Q^{\alpha b}\epsilon_{ba}\right){\bf P}^a$&$4\left({}^{16}t^c_\alpha ~{}^{45}\tau^{c}+{}^{16}\tau^c ~{}^{45}t^{c}_\alpha-\epsilon_{\alpha\beta\gamma} {}^{16}Q^{\alpha a} ~{}^{45}Q^{\gamma b}\epsilon_{ab}\right){\bf P}^\alpha$ \\

$J_3$&$\epsilon_{ijklm}\widehat{\bf M}^{ij}\widehat
{\bf P}^{kl}_{n}\widehat
{\bf F}^{mn}$&$\frac{1}{\sqrt{2}}f^{(45)}$&$\frac{8}{3}\left(~{}^{16}t^c_\alpha ~{}^{45}Q^{\alpha b}\epsilon_{ba}+~{}^{45}t^c_\alpha ~{}^{16}Q^{\alpha b}\epsilon_{ba}\right)\widetilde{\bf P}^a$&$4\left(~{}^{16}\tau^c ~{}^{45}t^{c}_\alpha-~{}^{16}t^c_\alpha~{}^{45}\tau^{c}\right)\widetilde{\bf P}^\alpha$ \\

$K_2$&$\widehat{\bf M}^{ij}\widehat
{\bf Q}_{j}\widehat
{\bf F}^{i}$&$-\frac{1}{2\sqrt{10}}f^{(10)}$&$\left(~{}^{16}Q^{a\alpha}~{}^{10}b^c_\alpha -~{}^{16}\tau^c ~{}^{10}L_{b}\epsilon_{ba}\right){\bf Q}_a$&$\left(\epsilon^{\alpha\beta\gamma}~{}^{16}t^c_\beta ~{}^{10}b^{c}_\gamma+~{}^{16}Q^{a\alpha}~{}^{10}L_a\right){\bf Q}_\alpha$ \\

$K_3$&$\widehat{\bf M}^{ij}\widehat
{\bf Q}^k_{ij}\widehat
{\bf F}^{k}$&$\frac{1}{2\sqrt{2}}f^{(10)}$&$\left(2~{}^{16}\tau^c ~{}^{10}L_{b}\epsilon_{ba}-\frac{2}{3}~{}^{16}Q^{a\alpha}~{}^{10}b^c_\alpha\right)\widetilde{\bf Q}^a$&$\left(\epsilon^{\alpha\beta\gamma}~{}^{16}t^c_\beta ~{}^{10}b^{c}_\gamma-~{}^{16}Q^{a\alpha}~{}^{10}L_a\right)\widetilde{\bf Q}_\alpha$ \\

\end{tabular}
\end{ruledtabular}
\end{table*}

From tables \ref{tab:table1} and \ref{tab:table2} we can easily write out the mass matrix.
The matrices that used to diagonalize the lower right $3\times 3$ part of mass matrix Eq.(\ref{dmatrix}) are given by
\begin{eqnarray}
&&U_{b(t,\tau)}=\\
&&\begin{pmatrix}
1&0&0&0&0\\
0&1&0&0&0\\
0&0&\cos\theta_{U_{b(t,\tau)}}&-\sin\theta_{U_{b(t,\tau)}}&0\\
0&0&\sin\theta_{U_{b(t,\tau)}}&\cos\theta_{U_{b(t,\tau)}}&0\\
0&0&0&0&1\nonumber
\end{pmatrix},
\end{eqnarray}
and
\begin{eqnarray}
&&V_{b(t,\tau)}=\\
&&\begin{pmatrix}
1&0&0&0&0\\
0&1&0&0&0\\
0&0&\cos\theta_{V_{b(t,\tau)}}&-\sin\theta_{V_{b(t,\tau)}}&0\\
0&0&\sin\theta_{V_{b(t,\tau)}}&\cos\theta_{V_{b(t,\tau)}}&0\\
0&0&0&0&1\nonumber
\end{pmatrix}.
\end{eqnarray}
They are just the matrices used in Ref.\cite{Babu1}, when eliminating the upper-left $2\times2$ identity matrices.

The matrices used for $t$ and $\tau$ have the same form. All the matrices elements are given by\cite{Babu1}
\begin{equation}\label{Bangle}
\tan\theta_{U_b}=-\frac{f^{(10)}q}{\sqrt{2}m_f^{(10)}}, ~\tan\theta_{V_b}=\frac{\sqrt{2}f^{(45)}p}{m_f^{(45)}},
\end{equation}
\begin{equation}
\tan\theta_{U_\tau}=\frac{6\sqrt{2}f^{(45)}p}{m_f^{(45)}},~\tan\theta_{V_\tau}=\frac{3\sqrt{2}f^{(10)}q}{4m_f^{(10)}},
\end{equation}
and
\begin{equation}
\tan\theta_{U_t}=-\frac{4\sqrt{2}f^{(45)}q}{m_f^{(45)}}, ~\tan\theta_{V_t}=\frac{\sqrt{2}f^{(45)}p}{m_f^{(45)}}.
\end{equation}

When taking the quartic couplings for up type quarks to be
diagonalized, we can easily get the Yukawa coupling constants for masses.
\begin{eqnarray}\label{hu}
h_u&=&\Big[\frac{24p}{\sqrt{5}}\xi^{(10)}+\frac{6p}{\sqrt{5}}\left(-\lambda^{(45)}+\frac{4}{3}\xi^{(120)}\right)\Big]\cos\theta_D\nonumber\\
&&-\sqrt{6}p\Big[20\xi^{(10)}-\frac{7}{12}\lambda^{(45)}-\frac{9}{4}\lambda^{(54)}\nonumber\\
&&+\frac{20}{9}\xi^{(120)}\Big]\sin\theta_D,
\end{eqnarray}
\begin{eqnarray}
f_d&=&\frac{q\cos\theta_D}{\sqrt{5}}\left(\frac{11}{15}\varrho^{(126)}+8\zeta^{(10)}-\lambda^{(10)}+\frac{56}{3}\zeta^{(120)}\right)\nonumber\\
&&+2\sqrt{6}q\left(\frac{1}{9}\varrho^{(126)}+8\zeta^{(10)}-\frac{32}{9}\zeta^{(120)}\right)\sin\theta_D,\nonumber\\
\end{eqnarray}
\begin{eqnarray}
f_e&=&\frac{3q\cos\theta_D}{\sqrt{5}}\left(-\frac{3}{15}\varrho^{(126)}+8\zeta^{(10)}-\lambda^{(10)}-\frac{8}{3}\zeta^{(120)}\right)\nonumber\\
&&+2\sqrt{6}q\left(\frac{1}{3}\varrho^{(126)}-8\zeta^{(10)}-\frac{4}{3}\zeta^{(120)}\right)\sin\theta_D,\nonumber\\
\end{eqnarray}
\begin{eqnarray}
h_t&=&3\sin\theta_{U_t}\cos\theta_{V_t}f^{(45)}\left(\frac{\cos\theta_D}{\sqrt{10}}+\frac{\sin\theta_D}{2\sqrt{3}}\right),\nonumber\\
\end{eqnarray}
\begin{eqnarray}
f_b&=&-\frac{1}{2}\sin\theta_{U_b}\cos\theta_{V_b}f^{(10)}\left(\frac{\cos\theta_D}{\sqrt{10}}+\frac{\sin\theta_D}{2\sqrt{3}}\right),\nonumber\\
\end{eqnarray}
and
\begin{eqnarray}
f_\tau&=&-\frac{1}{2}\sin\theta_{U_\tau}\cos\theta_{V_\tau}f^{(10)}\left(-\frac{\cos\theta_D}{\sqrt{10}}+\frac{\sqrt{3}\sin\theta_D}{2}\right).\nonumber\\
\end{eqnarray}
We have neglected the generation indices in $h_u$, $f_d$ and $f_e$, which should be $3\times3$ matrices. $h_t$, $f_b$ and $f_\tau$ are the Yukawa couplings after diagonalization of cubic coupling matrices.

The couplings for Baryon-Lepton number violating terms $QQ$'s and $QL$'s used in section \ref{sec:bl} are given by
\begin{eqnarray}
Y_{Q1}&=&-32q\left(\frac{2}{15}\varrho^{(126)}+\zeta^{(10)}+\frac{2}{3}\zeta^{(120)}\right),\\
Y_{Q2}&=&\frac{16p}{\sqrt{5}}\xi^{(10)}+\frac{4p}{\sqrt{5}}\left(-\lambda^{(45)}+\frac{4}{3}\xi^{(120)}\right),\\
Y_{Q3}&=&5\sqrt{2}p\left(3\lambda^{(54)}+\lambda^{(45)}-\frac{16}{3}\xi^{(120)}-4\xi^{(10)}\right),\nonumber\\
\end{eqnarray}
\begin{eqnarray}
Y_{L1}&=&2p(-\lambda^{(45)}+5\lambda^{(54)}-16\xi^{(10)}-\frac{16}{3}\xi^{(120)}),\\
Y_{L2}&=&\frac{q}{\sqrt{5}}\left(-\frac{14}{15}\varrho^{(126)}+3\lambda^{(10)}-16\zeta^{(10)}-\frac{64}{3}\zeta^{(120)} \right),\nonumber\\ \\
Y_{L3}&=&\sqrt{2}q\left(40\zeta^{(10)}-\frac{1}{15}\varrho^{(126)}-\frac{16}{3}\zeta^{(120)}\right),
\end{eqnarray}
and
\begin{eqnarray}
Y_{Q2}^{33}&=&\frac{4f^{(45)}}{\sqrt{10}}\cos\theta_{V_t}\sin\theta_{V_t}.
\end{eqnarray}
Except for the last one, the other equations carry generation indices and are matrices.
It is important to note that after taking quartic couplings for the up quark to be diagonal, $Y_Q$'s and $Y_{L1}$ have only diagonal elements and $Y_{L1}$ has no contribution to Eq.(\ref{fl}).

\section{Fields Normalization}
In deducing table \ref{tab:table1} and \ref{tab:table2}, we have used
\begin{equation}
\partial\widehat{\bf P}^{ij}_k\partial\widehat{\bf P}^{ij}_k=\partial\widetilde{\bf P}^a\partial\widetilde{\bf P}^a+\partial\widetilde{\bf P}^\alpha\partial\widetilde{\bf
P}^\alpha+\cdots.
\end{equation}
So the triplets and doublets are normalized according to
\begin{eqnarray}
(\partial\widetilde{\bf Q}_a,\partial\widetilde{\bf P}^a)&\to&\frac{\sqrt{3}}{2\sqrt{2}}(\partial\widetilde{\bf Q}_a,\partial\widetilde{\bf P}^a),\nonumber\\
(\partial\widetilde{\bf Q}_\alpha,\partial\widetilde{\bf P}^\alpha)&\to&\frac{\sqrt{2}}{2}(\partial\widetilde{\bf Q}_\alpha,\partial\widetilde{\bf P}^\alpha).
\end{eqnarray}

The other fields' normalizations can be found in \cite{Babu1,Babu}.

\bibliography{fmpd}

\end{document}